\documentclass[12pt]{iopart}

\usepackage{graphicx}
\usepackage[german,english]{babel}
\usepackage{iopams}
\usepackage{subfigure}
\usepackage{braket}

\begin{document}

\title[Effects of valence, geometry and electronic correlations on transport]{Effects of valence, geometry and electronic correlations on transport in transition metal benzene sandwich molecules}

\author{M. Karolak}
\address{Institut f{\"u}r Theoretische Physik und Astrophysik, Universit\"{a}t W\"{u}rzburg, Am Hubland, 97074 W\"{u}rzburg, Germany}

\author{D. Jacob}
\address{Max-Planck-Institut f{\"u}r Mikrostrukturphysik, Weinberg 2, 
06120 Halle, Germany}

\ead{mkarolak@physik.uni-wuerzburg.de}

\begin{abstract}

We study the impact of the valence and the geometry on the electronic structure and
transport properties of different transition metal-benzene sandwich molecules bridging 
the tips of a Cu nanocontact. Our density-functional calculations show that the electronic transport properties of the molecules depend strongly on the molecular geometry which can be controlled by the nanocontact tips. Depending on the valence of the transition metal center certain molecules can be tuned in and out of half-metallic behaviour facilitating potential spintronics applications.
We also discuss our results in the framework of an Anderson impurity model, indicating cases where the inclusion of local correlations alters the ground state qualitatively. For Co and V centered molecules we find indications of an orbital Kondo effect.
                                                                                                                                                                                                                                      
\end{abstract}


\maketitle

\section{Introduction}
\label{introduction}
Nanoscale devices such as atomic and molecular conductors offer an experimental handle to control the electronic and transport properties of these systems 
by manipulation of the molecular geometry \cite{iancu,choi}. Particularly interesting systems both from the point of view of application and
fundamental physics are magnetic molecules deposited either on metal surfaces 
or contacted by metallic contacts. From the point of view of application such 
devices built from molecular magnets\cite{gatteschi} offer the possibilty of 
ultimately miniaturized magnetic storage devices and/or 
for spintronics applications \cite{wolf2001,zutic2004,bogani2008,schmaus2011}.

On the other hand, whenever a magnetic atom or molecule is coupled to a metal substrate 
or metal electrodes the Kondo effect can arise \cite{kondo,hewson}.
Usually, the Kondo effect leads to the screening of the magnetic moment by 
the conduction electrons of the metal due to the formation of a total spin-singlet state,
and is signalled by the appearance of a sharp and strongly temperature-dependent resonance 
in the spectral function at the Fermi level, the so-called Kondo resonance or 
Abrikosov-Suhl resonance. The Kondo resonance in turn gives rise to a zero-bias anomaly 
in the conductance characteristics of the nanoscale device. In fact, such zero bias anomalies 
have been observed in numerous experiments involving magnetic atoms and molecules depositied on 
surfaces or attached to leads \cite{Madhavan1998s,Berndt_98,Knorr_2002,zhao2005,Yu2005,gao2007,Neel_2010,novel_kondo,jacob_2013,Kuegel_MnPc,tpp-paper}.

Here we study the very simple molecular magnets, formed by one transition metal ion from the $3d$ series (Sc,Ti,V,Co,Ni) between two benzene (C$_6$H$_6$) rings, trapped in a Cu nanocontact. We calculate the densities of states as well as transmission functions using the Landauer formula for different electrode separations. We find that a manipulation of the molecular geometry by applying pressure with the nanocontact tips can change the low-bias transport properties of the molecules qualitatively from insulating to metallic, via a half-metallic magnetic phase. This indicates that these molecules could be used in spintronics applications. An analysis of the local electronic structure of the central ion in terms of an Anderson impurity model gives a hint towards the expected electronic correlation effects in these systems. In a paper of ours \cite{sandwich_prl} we demonstrated that electronic correlations can lead to the occurence of the Kondo effect in the CoBz$_2$ system. Here, we show that the same seems to 
occur in VBz$_2$ albeit in a different set of orbitals.

% not attempt to 
%calculate the spectra including correlation effects, but will discuss possible candidates apart from CoBz$_2$ in the framework of an Anderson model, where correlation effects could play an important role.

\section{Methodology}

The now standard approach for calculating the electronic structure and transport properties of nanoscale 
conductors consists in combining the Kohn-Sham density functional theory (DFT) calculations with 
the Landauer or non-equilibrium Greens function approach\cite{palacios2001,taylor2001}.
In this approach the Kohn-Sham DFT effectively yields a (static) mean-field approximation
for the complicated many-body problem. 

Here we will employ the DFT based transport approach for nanoscopic conductors implemented in the 
ALACANT software package\cite{alacant,jacob2011}. This package has recently been
extended by one of us\cite{jacob2009,jacob2010_1,jacob2010_2,davids_review} in order to capture the 
effect of dynamic correlations arising from strong local interactions by adapting the
DFT++ approach which is the de facto standard in the theory of solids\cite{kotliar_rmp,lda++}
to the case of nanoscopic conductors. In this approach the incorrect behaviour of the Kohn-Sham 
DFT for strongly correlated electrons is remedied by augmenting the DFT with a local Hubbard-like 
interaction. We will limit ourselves to the DFT based transport calculations, however, since the required solution of the Anderson impurity model is very time consuming.

For the DFT calculations we use the CRYSTAL06 code\cite{crystal} employing the LDA\cite{kohn_sham}, 
PW91\cite{pw91} and the hybrid functional B3LYP\cite{becke}, together with the all electron Gaussian 
6-31G basis set. The geometries of the wires were relaxed beforehand and kept fixed during the calculations. 
The geometry of the molecule in contact with the wires was relaxed employing the B3LYP functional.

\begin{figure}
  \begin{center}
    \includegraphics[width=0.75\linewidth]{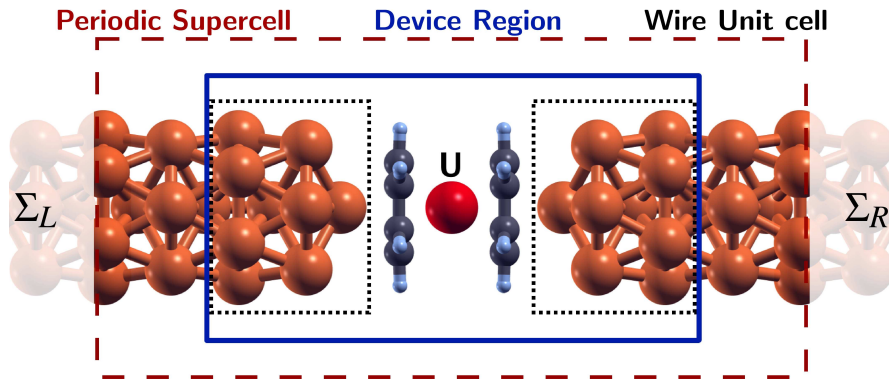}
  \end{center}
  \caption{Schematic picture showing the division of the system into device region D
    and the two leads L and R.  
    The periodic supercell as well as the wire unit cells are used in DFT 
    calculations of the system. The device region is the primary scattering region 
    in the subsequent transport calculations. The lead self energies representing the 
    coupling of the device region to the semi-infinite leads are indicated by $\Sigma_L$ 
    and $\Sigma_R$, a possible local Coulomb interaction on the central atom is indicated by $U$.}
  \label{method_1}
\end{figure}

In order to perform the DFT electronic structure calculations, the system is divided into three parts: 
The two semi-infinite leads L and R and the device region D containing the molecule and part of the 
leads. The schematic in Fig.{\ref{method_1}} shows the different regions used here. The calculations on 
the DFT level commence as follows: First, a calculation with the periodic unit cell shown in Fig.{\ref{method_1}} 
is performed. Additionally, the infinite wire is calculated using both left and right unit cells. Then the device 
region is cut out and the semi-infinite leads are attached on both sides. The Kohn-Sham Green's function of region 
the device region D can now be obtained from the DFT electronic structure as
\begin{equation}
  \label{eq:G0}
  G_{\rm D}(\omega)=(\omega+\mu-H^0_{\rm D}-\Sigma_{\rm L}(\omega)-\Sigma_{\rm R}(\omega))^{-1}
\end{equation}
where $H^0_{\rm D}$ is the Kohn-Sham Hamiltonian of region D and $\Sigma_{\rm L,R}(\omega)$ are 
the so-called lead self-energies which describe the coupling of the device region to L and R 
and which are obtained from the DFT electronic structure of the nanowire leads.

From the Green function $G_{\rm D}$ we can calculate the transmission 
function which describes the coherent transport through the device:
\begin{equation}
  \label{eq:Transm}
  T(\omega)={\rm Tr}[\Gamma_{\rm L}G_{\rm D}^\dagger\Gamma_{\rm R}G_{\rm D}]
\end{equation}
where $\Gamma_{\alpha}\equiv i(\Sigma_{\alpha}-\Sigma_{\alpha}^\dagger)$ and $\alpha = \rm L, \rm R$. 
For small bias voltages $V$, the transmission yields the conductance: $\mathcal{G}(V)=(2e^2/h)T(eV)$.

\begin{figure}
  \begin{center}
    \includegraphics[width=0.5\linewidth]{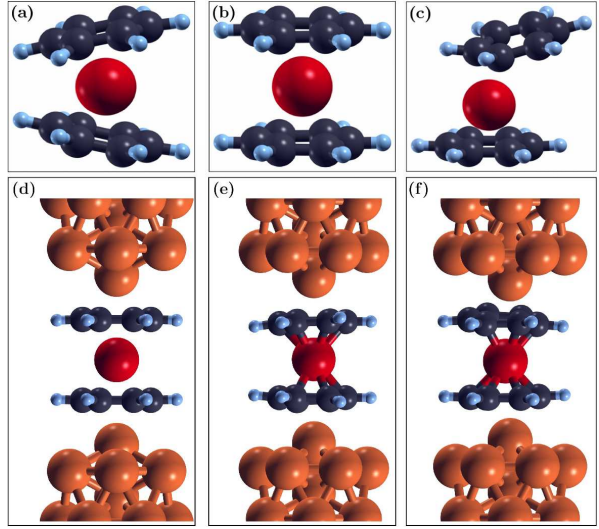}
  \end{center}
  \caption{\label{molecule_geom}
    (Color online) The three structures found in the literature for TMBz$_2$ molecules: 
    (a) the so-called onion structure, (b) the symmetric sandwich structure and (c) the asymmetric 
    sandwich structure. Relaxed geometries of CoBz$_2$ between Cu nanowires obtained starting 
    from a distance $d=3.6$~\r{A} between Co atom and Cu tip atoms and (d) a linear geometry, 
    (e) a stronlgy tilted geometry, (onion state). (f) Geometry obtained with the same starting molecule 
    structure as in (b) but with a smaller electrode separation $d=3.4$~\r{A}.
  }
\end{figure}
The $3d$ shell of the transition metal coupled to the rest of the system (benzene+leads) 
can be viewed as a so-called generalized Anderson impurity model (AIM)\cite{anderson_61}.
The AIM describes a situation where a magnetic impurity with strong local electronic correlations is embedded or more generally coupled to a bath of conduction electrons that are assumed to be non-interacting. It can be written in the multi-orbital case as

\begin{eqnarray}
\hat{H}_{\rm AIM}&=&\sum_{\nu}\varepsilon^{\phantom{\dagger}}_{\nu} \hat{c}^\dagger_{\nu}\hat{c}^{\phantom{\dagger}}_{\nu}
-\mu\sum_i \hat{d}^\dagger_{i}\hat{d}^{\phantom{\dagger}}_i+\sum_{\nu i}\left(V^{\phantom{\ast}}_{\nu i} \hat{c}^\dagger_{\nu}\hat{d}^{\phantom{\dagger}}_i+V_{\nu i}^\ast \hat{d}^\dagger_i \hat{c}^{\phantom{\dagger}}_{\nu}\right) \nonumber \\
&&\underbrace{+\sum_{i} \varepsilon^{\phantom{\dagger}}_{i} \hat{d}^\dagger_{i}\hat{d}^{\phantom{\dagger}}_i+\frac{1}{2}\sum_{ijkl}U^{\phantom{\dagger}}_{ijkl}\hat{d}^\dagger_{i}\hat{d}^\dagger_{j}\hat{d}^{\phantom{\dagger}}_l \hat{d}^{\phantom{\dagger}}_k}_{\hat{H}_{\rm loc}},
\label{am_hamilt}
\end{eqnarray}

where $\hat{c}_{\nu}$ and $\hat{d}_{i}$ are the bath and impurity degrees of freedom respectively. Here we identify the transition-metal center of the molecules as the impurity and absorb the rest of the system into the bath, which is assumed to be well described by DFT. We have also included a static shift of the impurity versus the bath given by $\mu$ along with the static crystal field $\varepsilon_i$ and the full local Coulomb interaction $U_{ijkl}$ in the last two terms. The energy levels of the bath are given by $\varepsilon_{\nu}$ and the hybridization parameters $V_{\nu i}$ give the amplitude for transitions of particles from the bath onto the impurity and vice versa. 
The model is completely defined by the Coulomb interaction parameters $U$ and $J$, the energy levels $\epsilon_i$ of the $3d$ 
orbitals and the so-called hybridization function $\Delta_i(\omega)$ (we assume a diagonal hybridization function).
The latter describes the (dynamic) coupling of the transition metal $3d$-shell to the rest of the system and is obtained by integrating out the bath degrees of freedom to yield
\begin{equation*}
\Delta_{i}(\omega)=\sum_{\nu}\frac{ V^{\phantom{\ast}}_{\nu i} V_{\nu i}^\ast}{\omega-\varepsilon_{\nu}+i\delta}.
\end{equation*}
It can equivalently be obtained from the Kohn-Sham Green's function as 
\begin{equation}
  \label{eq:Delta}
  \Delta_i(\omega)=\omega+\mu-\epsilon_i-[G^0_i(\omega)]^{-1},
\end{equation}
where $\mu$ is the chemical potential, $\epsilon_i$ are the Kohn-Sham energy levels 
of the $3d$-orbitals and $G^0_i(\omega)$ is the Kohn-Sham Green's function projected 
onto the $3d$ subspace. To solve the local impurity problem we use the one-crossing approximation (OCA) \cite{haule2001,kotliar06}. Within this implementation the Coulomb interaction is treated approximately including density-density terms $\propto \hat{n}_{i\sigma}=\hat{d}^\dagger_{i\sigma}\hat{d}^{\phantom{\dagger}}_{i\sigma}$, as well as the spin-flip term of the Kanamori Hamiltonian \cite{kanamori63}. In this approximation the Coulomb interaction part of the local Hamiltonian can be written as

\begin{equation*}
\hat{H}^{\rm OCA}_{\rm ee}=\frac{1}{2}\sum_{i,\sigma}U \hat{n}_{i,\sigma}\hat{n}_{i,-\sigma} 
+\frac{1}{2}\sum_{i\neq j,\sigma,\sigma^\prime}\left[\left(U-\frac{5J}{2}\right)\hat{n}_{i\sigma}\hat{n}_{j\sigma^\prime}\right]-\sum_{i\neq j}J\hat{\underline{S}}_i\hat{\underline{S}}_j,
\end{equation*}
with the spin-vector-operators $\hat{\underline{S}}_i$. Here we neglect the pair-hopping term \cite{kanamori63} and use a simplified interaction which only takes into account the direct Coulomb repulsion $U\equiv U_{ijij}$ and the Hund's rule coupling $J\equiv U_{ijji}$. We assume that the interaction is somewhat increased as compared to the bulk values since the screening should be weaker than in bulk. Ab initio values for the bulk obtained by the constrained random phase approximation (cRPA) have been reported, e.g., in Ref. \cite{ersoy_2011}. The values used here will be provided below; but indeed we find that our results are qualitatively stable for a reasonable range of values for $U$ and $J$ (see below). An estimation based on the work by Solovyev et al. \cite{solovyev_1996} performed in Ref. \cite{weng_2008_1} for TiBz and VBz clusters indicates also a value of $U\sim 3$eV for these molecules. We note that the method presented in Ref. \cite{solovyev_1996} only the $t_{2g}$ states are assumed to be 
correlated, while the $e_g$ states are treated as itinerant and contribute to the screening. 

As usual in DFT++ approaches a double counting correction (DCC) has to be subtracted to compensate for the overcounting of interaction terms. In the calculations involving Co and Ni we employ the so-called fully localized (FLL) or atomic limit correction \cite{amf_fll}
\begin{equation}
  \label{eq:dc}
  \mu_{\rm DC}=U\left(N_{\rm 3d}-\frac{1}{2}\right)-J\left(\frac{N_{\rm 3d}}{2}-\frac{1}{2}\right).
\end{equation}
This approach was successfully applied to single $3d$ transition metal atoms (Fe, Co, Ni) in nanocontact junctions \cite{jacob2009}. We find that this correction works well also for the molecules with the late transition metals Co and Ni as centers. However, we have found that the FLL correction leads to unrealistically high occupancies when applied to molecules centered on the early transition metals Sc, Ti, V. The same would be true on an even larger scale for the AMF correction. As is known already from DFT+U and related approaches, that the double counting can not be rigorously defined. For the early transition metal sandwiches we have thus adjusted the occupation of the $3d$ shell to be close to the value obtained from DFT. The DCC can be viewed as an impurity chemical potential and gives one the freedom to shift the impurity levels around to a certain degree. This can be exploited to mimic the effect of an additional 
gate electrode on the nanosystem and to push the system into different regimes.

The Anderson impurity model was introduced for the description of $d$ or $f$ shell impurities in simple metal hosts and is capable of describing local correlation physics, like the Kondo effect \cite{Schrieffer_Wolff,hewson}. An analysis of the hybridization function and interaction can already give a strong hint towards the expected many-body behaviour in the system, since they, enter as the determining factors in the estimate of the Kondo temperature.
In a simple model of the Kondo effect, i.e.\ the one-band case with a flat bath\cite{haldane,hewson}, the Kondo temperature $T_{\mathrm{K}}$ is given by the following formula:
\begin{equation}
k_{\mathrm{B}}\,T_{\mathrm{K}} = \frac{\sqrt{\Gamma_d U}}{2} \exp\left(\frac{\pi\varepsilon_d (\varepsilon_d+U)}{\Gamma_d U}\right), \label{kondo_equation}
\end{equation}
where $U$ is the Coulomb repulsion energy between single and double occupied state, 
$\varepsilon_d$ is the energy level of the single occupied state, 
and $\Gamma_d$ is the (constant) substrate-induced broadening of these states, proportional to the hybridization function.

\section{Transition-Metal Benzene Sandwich Molecules}
\label{general_obs}

\begin{figure}
  \begin{center}
    \includegraphics[width=0.6\linewidth]{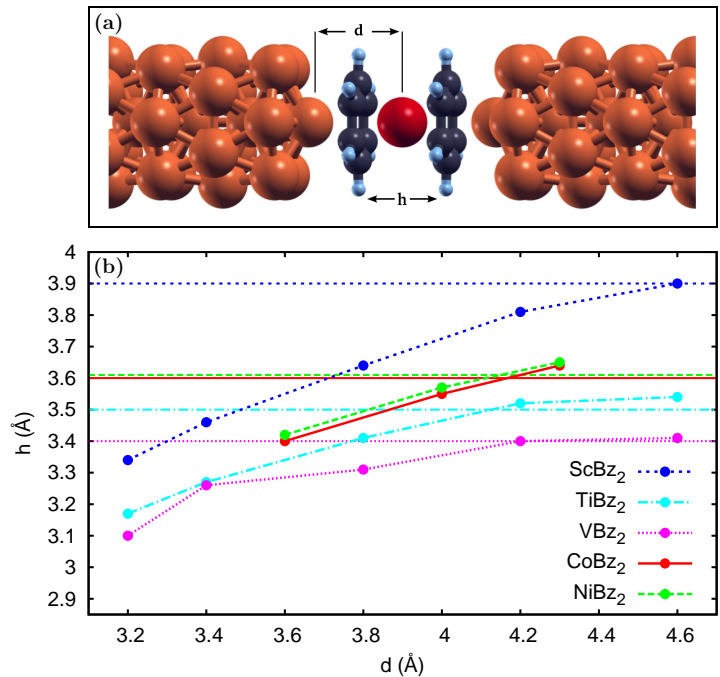}
  \end{center}
  \caption{(Color online) (a) The geometry used in the nanocontact calculations along with definitions of the distances $d$ and $h$. 
    (b) Dependence of the molecule size on the electrode to TM center separation. Equilibrium sizes of the free molecules are indicated 
    by corresponding dashed horizontal lines in the same linestyle and color.
    \label{general1}
  }
\end{figure}

We consider single transition-metal benzene (TMBz$_2$) molecules in contact with two semi-infinite Cu 
nanowires as shown schematically in Fig. \ref{general1}a. The TMBz$_2$ molecules are the smallest 
instance of a general class of M$_n$Bz$_{n+1}$ complexes, where M stands for a metal atom, 
that have been prepared and investigated starting from the prototype CrBz$_2$ in 1955, see Refs. \cite{silverthorn_1975,muetterties_1982} for reviews. 
Vanadium half sandwich VBz complexes were first synthesized and characterized in pentane solution by Andrews and Ozin in the mid 1980s. These discoveries along with extensive experimental and theoretical calculations also for VBz$_2$ (that we will revisit below) on the level of $X_\alpha$ theory were reported in a series of papers \cite{andrews_1986_1,andrews_1986_2}. 

Starting about a decade later Kaya and colleagues \cite{hoshino_1995,kurikawa_1995,kurikawa1999,nakajima2000} have synthesized and investigated general metal-benzene complexes comprised of $n$ metal atoms and $m$ benzene rings: M$_n$Bz$_m$. They successfully produced a variety of complexes in the gas phase using the whole $3d$ transition metal series from Sc to Cu. Concerning the structure of the complexes with M$=$Sc,Ti,V the authors established, experimentally, that the complexes form linear sandwich clusters of the form M$_n$Bz$_{n+1}$ without exterior metal atoms, whereas the late transition metals form so-called rice ball structures where one or more metal atoms are covered by benzene rings in an almost spherical fashion. Stern-Gerlach experiments for the early transition metal sandwiches revealed a magnetic moment of $0.4\mu_{\rm B}$ ($0.4$ Bohr magnetons) for ScBz$_2$ and $0.7\mu_{\rm B}$ for VBz$_2$ and no magnetic moment for TiBz$_2$ \cite{miyajima_2007}.

On the theory side these complexes were studied using different methods ranging from simple ligand field calculations over $X_\alpha$ and DFT calculations up to wave function based quantum Monte Carlo by various authors usually focussing on a specific metal center.  
The generic bonding mechanism for the MBz$_2$ complexes was established from ligand field theory arguments supported by $X_\alpha$ calculations for CrBz$_2$ \cite{weber_1978, osborne_1987} and VBz$_2$ \cite{andrews_1986_2}. The mechanism is similar to the bonding of single transition metal atoms to a single benzene ring \cite{bauschlicher_1992}. First, ligand field theory for the $C_{6v}$ symmetry dictates that the $3d$ orbitals of the metal center will be split into three irreducible representations: $a^\sigma_{1g}(d_{z^2})$, $e^\pi_{1g}(d_{xz},d_{yz})$ and $e^\delta_{2g}(d_{xy},d_{x^2-y^2})$. Notations concerning the symmetry differ in the literature, the representations sometimes being called $a_1,e_1,e_2$ or just using the symbol for the symmetry group and representation synonymously, capitalizing the symbol $A_1,E_1,E_2$. In the results section we will use the latter notation. The $z$-axis is here the axis going through the center of the benzene rings and constitutes the 6-fold rotation axis. The $e^\delta_{2g}$ orbital has a strongly bonding character through in-phase combinations of the in-plane metal orbitals with the benzene $\pi$ orbitals, leading to its energetic stabilization, i.e. it is energetically most favorable and occupied preferentially. The $e^\pi_{1g}$ orbitals form out of phase combinations with the ligand $\pi$ electron system and are thus antibonding in character, while the $a_{1g}$ remains nonbonding since it points through the center of the benzene. The primary bonding interactions are brought about by electron transfer from the benzene $\pi$ to the metal $e^\pi_{1g}$ relieved by backtransfer from the metal $e^{\delta}_{2g}$ to the $\pi^\ast$ orbitals of the ligands. In $X_\alpha$ calculations by Andrews et al. \cite{andrews_1986_2} it could be shown that the attachment of benzene rings to a V atom leads to the loss of $0.5$ $3d$ electrons per benzene attached, leading to a $3d^4$ configuration on the metal. At the same time the direct interaction between the benzene rings is small 
due to their relative large distance. These mechanisms in principle apply to the whole series of TMBz$_2$ molecules. 

Following the successful preparation of TMBz$_2$ complexes in the gas phase the interest in these complexes was renewed. A DFT study for the whole series from Sc to Ni was carried out by Pandey et al. \cite{pandey_2001} who investigated TMBz and TMBz$_2$ complexes and their respective cationic and anionic versions. Assuming a hexagonal $D_{6h}$ symmetry (i.e. the form of the complex shown in Fig. \ref{molecule_geom} (b)) they find a nonmonotonic evolution of the size of the TMBz$_2$ complexes. By size, here, the distance between metal atom and center of the benzene rings is meant. The authors, however, note, that although their calculations agree quite well with experiments for Sc to Cr, the agreement is poor for Co and Ni complexes. They speculate that the assumption that the CoBz$_2$ and NiBz$_2$ complexes have a $D_{6h}$ symmetry might be incorrect, as is actually the case as we will see below. Pandey et al. calculate also the valence configurations and spin multiplicities for all complexes, assuming $D_{
6h}$ 
symmetry. Weng and co-workers \cite{weng_2008_1,weng_2008_2} performed GGA and for the first time also GGA+U calulations for the symmetric sandwich structures of Sc,Ti and V centered clusters with TM$_n$Bz$_{n+1}$ typically for $n=2$. The authors focussed on magnetic properties, especially on how they can be manipulated and tuned from FM to AFM order within the cluster. They found that an increasing local Coulomb interaction $U$ leads to a general increase of the magnetic moment in the Vanadium $3d$ shell in a finite cluster as well as for an infinite chain. 

More recent calculations for CoBz$_2$ \cite{zhang_2008} and NiBz$_2$ \cite{rao_2002} revealed that the ground state of these systems is an asymmetric sandwich arrangement with merely $C_1$ symmetry (see Fig. \ref{molecule_geom} (c)). The lowering of energy by breaking the $D_{6h}$ symmetry can be understood as an effect of the 18 electron rule \cite{lauher_76,crabtree_2009}, that causes the complexes with more than 18 valence electrons (CoBz$_2$, NiBz$_2$ have 21, 22 respectively) to bend due to the symmetry of the HOMO (highest occupied molecular orbital). The HOMO is lowered in energy when the relative orientation of the two benzene rings is tilted. In the asymmetric sandwich structure the magnetic moments were found to be $1\mu_{\rm B}$ (one Bohr magneton) for CoBz$_2$ and vanishing for NiBz$_2$. The quenching of the moment of Ni was explained by the strong hybridization between the Ni $3d$ states and the $\pi$ electrons of benzene \cite{rao_2002}. 

A recent study of the structural properties of TMBz and TMBz$_2$ with TM$=$Sc,$\ldots$,Zn within M{\o}ller-Plesset second order perturbation theory (MP2) calls the established DFT results into question \cite{youn_2012}. The authors find that the perfectly symmetric sandwich structure with $D_{6h}$ symmetry is formed only by VBz$_2$. CrBz$_2$ and MnBz$_2$ form sandwiches where the two benzene rings are slightly rotated around the $z$ axis (their respective $C_6$ axis) with respect to each other. The other complexes show either strongly distorted sandwich structures (Sc, Ti, Fe, Ni) or a completely different arrangement (Cu, Zn). The authors note in passing, that they were not able to obtain a geometry for CoBz$_2$ due to convergence problems. An ab initio wave function based QMC study could probably settle the matter of the structures, unfortunately it is computationally so demanding that at the moment only studies of TMBz half-sandwiches are possible and have been performed \cite{horvathova_2013}. Since 
the structures of TMBz show qualitatively the same structure for the whole $3d$ series the comparison between MP2 and QMC in that case can only give weak hints for TMBz$_2$. In addition the methods generally do not agree on the metal-benzene distances, MP2 giving values up to $0.5$\AA smaller than QMC. 

Considerable effort from theory was invested in the understanding of multidecker vanadium sandwich complexes \cite{wang2005,xiang2006,maslyuk_2006,mokrousov2007,weng_2008_2,horvathova_2012}, since they showed half-metallic behavior possibly allowing for spintronics applications. The V$_n$Bz$_{n+1}$ complexes or nanowires show a half-metallic ferromagnetic behavior, which means that one spin species shows weight at the Fermi level and is available for transport, while the other spin species is gapped and thus insulating. Such a situation is shown for example in Fig.\ref{dft-results-v} (b) for the $a^\sigma_1$ channel of VBz$_2$ in a nanocontact and could be exploited to build a spin filter device. Since we are not aware of any experimental confirmations of this behavior the discussion remains theoretical at this point. Additionally, it is unclear how a possible Kondo effect (see below) in such devices would influence the conductance characteristics of a general V$_n$Bz$_{n+1}$ device.

%%%%%%%%%%%%%%%%%
\section{DFT calculations}
%%%%%%%%%%%%%%%%%

\begin{figure}[t]
  \begin{center}
    \includegraphics[width=0.98\linewidth]{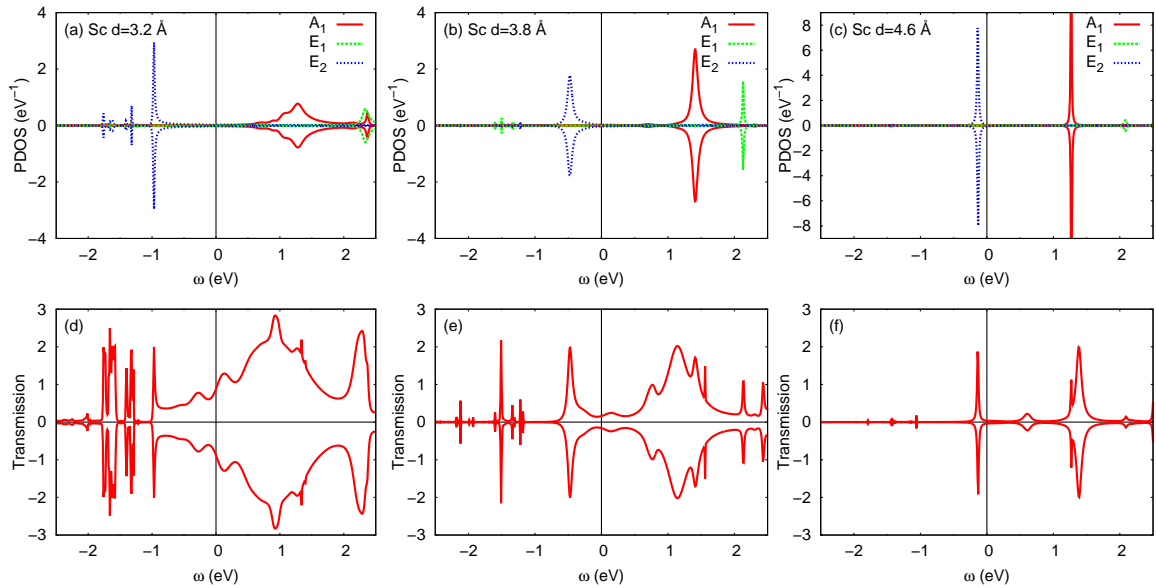}
  \end{center}
  \caption{
    (a) Projected density of states for the $3d$ shell of Sc for $d=3.2$~\r{A}, (b) $d=3.8$~\r{A} and (c) $d=4.6$~\r{A}. The corresponsing transmission functions are shown in panels (d), (e) and (f) respectively.
  }
  \label{dft-results-sc}
\end{figure}

When the molecules are brought in contact with electrodes, say in a break-junction or 
scanning tunneling microscope (STM) experiment, they will interact with the electrodes 
and their structure will be subject to change. We have used copper electrodes in a hexagonal 
geometry, as proposed in Ref. \cite{tosatti2001}. For ScBz$_2$, TiBz$_2$ and VBz$_2$, 
only small changes in geometry occur, the molecule remains in its symmetrical sandwich 
structure and can, in our idealized picture, only be compressed or elongated by the electrodes. 

Conversely, the asymmetric structures as formed by CoBz$_2$ and NiBz$_2$, show considerable changes in 
geometry when in contact with electrodes. In general though the benzene rings are brought into a more 
symmetric arrangement as shown in Fig.~\ref{molecule_geom}. Figure~\ref{molecule_geom}d shows the relaxed
geometry using $d=3.6$\AA \ and starting from the linear configuration of the CoBz$_2$ sandwich aligned 
with the axis defined by the Cu nanowires. In this case the linear geometry and alignment is preserved. 
Figure~\ref{molecule_geom}e shows the relaxed geometry obtained when starting from a 
strongly tilted geometry of the sandwich molecule (onion state). 

In this case the linear configuration of the idealized sandwich molecule is approximately recovered, 
but the molecule as a whole is slightly rotated with respect to the Cu nanowire axis. 
The lifting of the orbital degeneracy in the $E_2$ channel is on the order of few tens of meV.
We additionally show (Fig. \ref{molecule_geom}f) that the effect of the leads is even stronger 
at smaller lead-molecule separation. The relaxed geometry obtained for a distance $d=3.4$~\r{A} 
between the Co atom and the Cu tip atoms shows an even smaller tilting of the Benzene rings, leading to 
a reduced splitting in the $E_2$ channel.

Even starting from the strongly asymmetric sandwich structure the presence of the leads favours a parallel 
arrangement of the Bz rings. Thus in the end the molecules we investigated here are brought closer to the symmetric 
sandwich structure, when in contact with the leads. Hence for the present study we have assumed the symmetric sandwich structure for all molecules. 

We investigate the different molecules at different Cu-tip-TM distances. In the Case of Co and Ni we use 3.6~\r{A}, 
4.0~\r{A} and 4.3~\r{A} (see Fig.~\ref{general1}a,b), whereas for ScBz$_2$, TiBz$_2$ and VBz$_2$ we used 3.2~\r{A}, 
3.4~\r{A}, 3.8~\r{A}, 4.2~\r{A} and 4.6~\r{A}. The Bz-Bz distance $h$ varies depending on the distance $d$ of the 
Cu tip to the TM atom in the center of the molecule. Figure~\ref{general1}b shows the general trend of the molecule 
size versus the electrode distance, the free diameters (obtained for the symmetric sandwich structure) of the molecules 
are indicated also, showing that the molecules are compressed at small electrode separations. Horizontal lines with the 
same linestyle and color indicate the equilibrium geometries of the free molecules.

\begin{figure}[t]
  \begin{center}
    \includegraphics[width=0.98\linewidth]{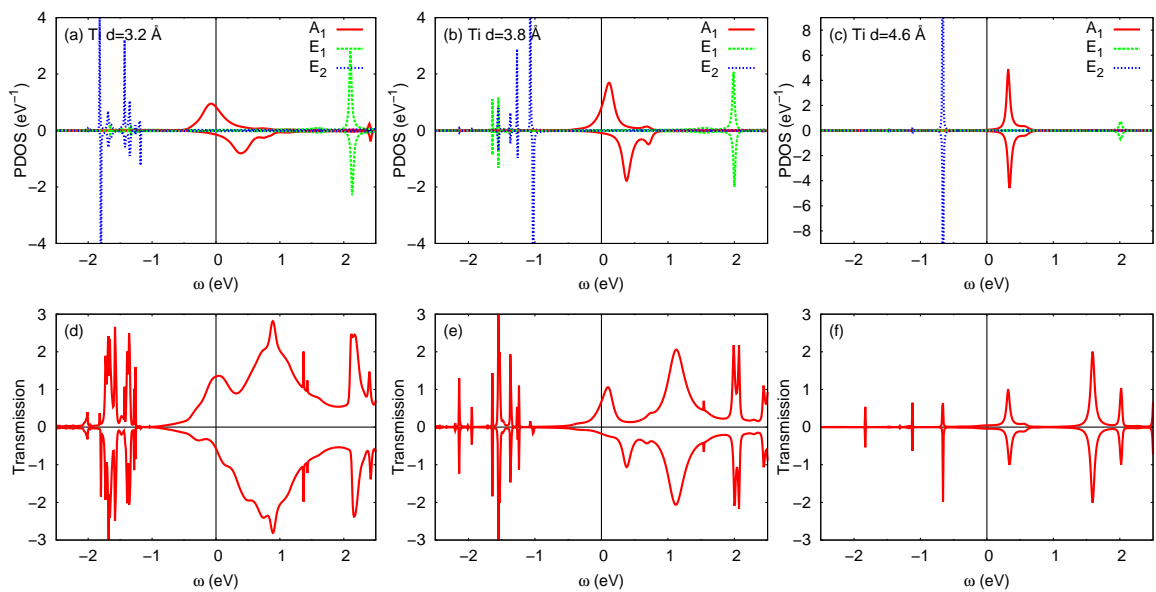}
  \end{center}
  \caption{
    (a) Projected density of states for the $3d$ shell of Ti for $d=3.2$~\r{A}, (b) $d=3.8$~\r{A} and (c) $d=4.6$~\r{A}. The corresponsing transmission functions are shown in panels (d), (e) and (f) respectively.
  }
  \label{dft-results-ti}
\end{figure}

We begin by discussing the electronic structure of the molecules in contact with the nanowires on the level of DFT. We have performed calculations using DFT with the L(S)DA and B3LYP functionals. We have collected data concerning occupancies in table \ref{lda-fillings} and magnetic moments in table \ref{mag_mom}. The LDA will show the smallest moment, while B3LYP contains a 20\% admixture of exact exchange and thus yields a moment hihger than LDA at all times.

As we mentioned already the hexagonal symmetry of the sandwich structure leads to a lifting of the degeneracy 
of the $3d$ shell. The strong crystal field splits the shell into a singlet $A_1$ consisting of the $d_{3z^2-r^2}$ and two doublets $E_1$ consisting of the $d_{xz}$ and $d_{yz}$ orbitals and $E_2$ consisting of the $d_{xy}$ and $d_{x^2-y^2}$ orbitals. We assume an atomistic point of view concerning the $3d$ shell of the central atom in the light of our subsequent Anderson impurity model treatment, i.e. we define the symmetry adapted basis in terms of the orbitals of the metal of the corresponding symmetry. Since the 6-31G basis set provides 10 (or 12 depending on the representation \cite{szabo_ostlund}) basis functions for the $d$ shell our five orbitals are formed by diagonalization of the 10 orbital set within the crystal field. The diagonalization yields in all cases five $3d$ orbitals in the vicinity of the Fermi level and five $3d$ orbitals very high $\sim 20-50$eV above the Fermi level. In this manner we obtain the best symmetry adapted atomic-like orbitals for our subsequent analysis.

\begin{figure}[t]
  \begin{center}
\includegraphics[width=0.98\linewidth]{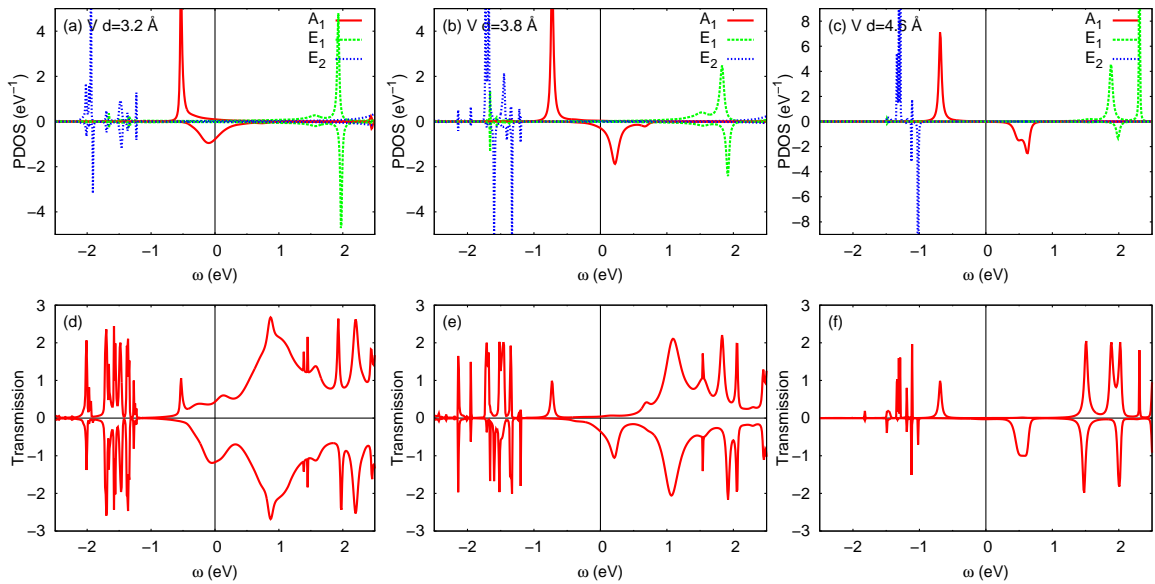}
  \end{center}
  \caption{
    (a) Projected density of states for the $3d$ shell of V for $d=3.2$~\r{A}, (b) $d=3.8$~\r{A} and (c) $d=4.6$~\r{A}. The corresponsing transmission functions are shown in panels (d), (e) and (f) respectively.
  }
  \label{dft-results-v}
\end{figure}

Let us begin with Sc. The projected densities of states (PDOS) of the $3d$ shell of Sc obtained from LSDA is shown in Fig. \ref{dft-results-sc} (a), (b) and (c) for increasing electrode-molecule separation. The two spin projections are plotted as the positive and negative ordinate respectively. Only a very small spectral weight of the $3d$ states at the Fermi level is observed in all cases. The changes over the whole range of $d$ are continuous and towards larger energy separations the levels become sharper showing the increasing decoupling of the molecule from the leads. The transmission function also follows this general trend. We want to point out that the transmission function reflects the transport through the whole molecule, not only the transition metal center. It is thus clear, that there are other states, like the benzene $\pi$ system contributing here, as one can see in Fig. \ref{dft-results-sc}(d), where the transport at low bias is certainly not brought about by transition metal $3d$ states. This 
effect is strongest at smallest electrode separations, because the benzene rings are forced closer together leading to direct interactions between their respective $\pi$ electron systems and to stronger interactions with the Cu electrodes' $s$ orbitals. For the smallest electrode-transition metal distance the Sc atom is occupied by two electrons equally distributed in the $E_1$ and $E_2$ shells, leaving the $A_1$ shell empty, see table \ref{lda-fillings}. Towards larger electrode separations the charge redistributes in favor of the $E_2$ set. Since the free Sc centered sandwich is the largest, its electronic structure is influenced considerably by the electrodes. One observes, for example, that the $E_2$ orbitals are energetically stabilized by the increased interaction with the $\pi$ electron system of the benzene rings if they are pushed closer towards the metal. Also the population of the $E_1$ set increases by the stronger interaction with the $\pi$ system. The system shows no magnetic polarization 
within LSDA, but within B3LYP at $d=4.6$~\r{A} or without electrodes a small magnetic moment of about 0.4$\mu_{\rm B}$ arises in the $3d$ shell, in accordance with Stern-Gerlach experiments reported in Ref. \cite{miyajima_2007}. Since the formal spin multiplicity $M=2S+1$ of the free molecule is 2, one unpaired electron is expected, its moment is probably quenched by the stronger interaction with the benzene rings at small electrode separations, similarly to NiBz$_2$ \cite{rao_2002}.

\begin{table}[t]%[H] add [H] placement to break table across pages
\caption{\label{mag_mom} Magnetic moments in units of the Bohr magneton $\mu_B$ of of the $3d$ shell of sandwich molecules in the Cu nanocontact using LSDA and B3LYP functionals.(\footnotemark[1] Symmetric Structure)}
 \begin{indented}
 \item[] \begin{tabular}{@{\extracolsep{\fill}} c c c c c c c}
 \br
 & 3.2~\r{A} & 3.4~\r{A}& 3.8~\r{A}& 4.2~\r{A} & 4.6~\r{A} & Free \\
 \mr
 Sc(LSDA) & 0.0 & 0.0 & 0.0 & 0.0 & 0.0 & \\
 Sc(B3LYP) & 0.0 & 0.0 & 0.0 & 0.0 & 0.25 & 0.37 \\
 \mr
 Ti(LSDA) & 0.45 & 0.44 & 0.31 & 0.15 & 0.0 & \\
 Ti(B3LYP) & 1.0 & 1.0 & 1.0 & 0.0 & 0.0 & 0.0 \\
  \mr
 V(LSDA) & 0.44 & 0.75 & 0.9 & 1.0 & 1.1 & \\
  V(B3LYP) & 1.15 & 1.25 & 1.25 & 1.30 & 1.32 & 1.45 \\
\mr
 \mr
 & &3.6~\r{A} & 4.0~\r{A}& 4.3~\r{A}&  &  Free \\
 \mr
 Co(LSDA) & & 1.2 & 1.15 & 1.4 & & \\
 Co(B3LYP) & & 2.0 & 2.0 & 2.0 & & 1.35 (1.44)\footnote[1]{}\\
 Ni(LSDA) & & 0.0 & 0.0 & 0.0 & & \\
 Ni(B3LYP) & & 0.0 & 0.0 & 0.0 & & 0.0\\
 \br
\end{tabular}
 \end{indented}
\end{table}

The case of Ti is different, as the system already shows a large $3d$ spectral weight at the Fermi level in LSDA at small electrode molecule separations, see Fig. \ref{dft-results-ti}(a) and (b). This weight is brought about by the $A_1$ channel, which compared to Sc now holds about one electron. The $A_1$ channel also dominates the transmission at small biasses as can be seen best in Fig. \ref{dft-results-ti}(e). At the smallest electrode separation states derived from benzene are pushed towards the Fermi level and contribute significantly to the transmission. The spin polarization of about 0.45$\mu_{\rm B}$ arises in LSDA when the molecule is in close contact with the leads at $d=3.2$~\r{A} and $d=3.4$~\r{A}. At $d=3.4$~\r{A} (not shown) the system is almost a half metal, one spin channel being conducting the other almost insulating, see Fig. \ref{dft-results-ti} (b), (e). This behavior makes it similar to the V$_n$Bz$_{n+1}$ systems, where this behavior has been predicted, see the discussion in section
\ref{general_obs}. At larger distances the magnetic moment continuously reduces and finally vanishes at $d=4.6$~\r{A} as shown in Fig. \ref{dft-results-ti} (c) and table \ref{mag_mom}. The occupancies shown in table \ref{lda-fillings} indicate that the $A_1$ channel gets depopulated at larger distances and thus the magnetic moment vanishes. B3LYP calculations show a similar behavior with a moment of 1 $\mu_{\rm B}$ at small separations and a vanishing moment at large separations and for the free molecule. It was shown in Stern-Gerlach experiments in Ref. \cite{miyajima_2007} that the free TiBz$_2$ molecule does not exhibit a magnetic moment, which is in accordance with our findings and the formally assigned spin multiplicity of 1, i.e. a singlet.

\begin{table}[t]%[H] add [H] placement to break table across pages
\caption{\label{lda-fillings} Fillings of the $3d$ shell on the LDA and for Co/VBz$_2$ also on the LDA+OCA level.}
 \begin{indented}
\item[] \begin{tabular}{@{\extracolsep{\fill}} c c c c c c c c c}
\br
&\multicolumn{4}{c}{{3.2~\r{A}}}&\multicolumn{4}{c}{{3.4~\r{A}}}\\
	& $A_1$ & $E_1$ & $E_2$ & Tot&$A_1$ & $E_1$ & $E_2$ & Tot\\
	\mr
	Sc (LDA) & $0.11$ & $0.87$ & $1.03$ & $2.00$ & $0.10$ & $0.85$ &  $1.04$ & $1.99$  \\
	Ti (LDA) & $0.45$ & $1.02$ &  $1.36$ & $2.83$ & $0.36$ & $1.00$ &  $1.39$ & $2.74$  \\
	V (LDA) & $1.33$ & $1.04$ &  $1.59$ & $3.97$ & $1.34$ & $0.98$ &  $1.63$ & $3.95$ \\
	V (LDA+OCA)& $0.89$ & $2.58$ & $0.61$ & $4.08$ & $0.90$ & $1.94$ &  $1.22$ & $4.06$ \\
 \mr
 \mr
&\multicolumn{4}{c}{{3.8~\r{A}}}&\multicolumn{4}{c}{{4.2~\r{A}}}\\
	& $A_1$ & $E_1$ & $E_2$ & Tot&$A_1$ & $E_1$ & $E_2$ & Tot\\
	\mr
	Sc (LDA) &  $0.09$ & $0.80$ &  $1.07$ & $1.96$ & $0.09$ & $0.75$ &  $1.15$ & $1.99$ \\
	Ti (LDA) & $0.24$ & $0.96$ &  $1.45$ & $2.65$& $0.17$ & $0.92$ &  $1.53$ & $2.61$ \\
	V (LDA) & $1.27$ & $0.97$ &  $1.68$ & $3.92$& $1.19$ & $0.93$ &  $1.76$ & $3.88$ \\
	V (LDA+OCA)& $0.91$ & $1.94$ &  $1.22$ & $4.07$ & $0.91$ & $1.93$ &  $1.22$ & $4.06$\\
 \mr
 \mr
&\multicolumn{4}{c}{{4.6~\r{A}}}&\multicolumn{4}{c}{}\\
	& $A_1$ & $E_1$ & $E_2$ & Tot\\
	\mr
	Sc (LDA) &  $0.09$ & $0.75$ &  $1.15$ & $1.99$ \\
	Ti (LDA) & $0.17$ & $0.92$ &  $1.53$ & $2.61$\\
	V (LDA) & $1.19$ & $0.93$ &  $1.76$ & $3.88$ \\
	V (LDA+OCA)& $0.91$ & $1.93$ &  $1.22$ & $4.06$\\
 \mr
 \mr
 &\multicolumn{4}{c}{{3.6~\r{A}}}&\multicolumn{4}{c}{{4.0~\r{A}}}\\
	&$A_1$ & $E_1$ & $E_2$ &Tot& $A_1$ & $E_1$ & $E_2$ &Tot \\
	\hline
	Co (LDA) & $1.78$ & $2.75$ &  $2.83$ & $7.36$ &  $1.80$ &  $3.52$ & $2.80$ & $7.85$  \\
	Co (LDA+OCA)& $1.02$ & $3.74$ &  $2.74$ & $7.50$ & $0.98$ &  $3.78$ & $3.28$ & $8.03$  \\
	Ni (LDA) & $1.82$ & $3.66$ &  $3.07$ & $8.55$ &  $1.80$ &  $3.69$ & $3.02$ & $8.51$ \\
\mr
 \mr
 &\multicolumn{4}{c}{{4.3~\r{A}}}&\multicolumn{4}{c}{}\\
	& $A_1$ & $E_1$ & $E_2$ & Tot \\
	\hline
	Co (LDA) &   $1.80$ & $3.54$ & $2.87$ & $8.21$ \\
	Co (LDA+OCA)& $0.96$ & $3.80$ & $3.50$ & $8.26$ \\
	Ni (LDA) &   $1.80$ & $3.74$ & $3.04$ & $8.58$\\

\br
\end{tabular}
 \end{indented}
\end{table}

The VBz$_2$ sandwich shows the converse behavior to the Ti centered sandwich. The magnetization increases for large $d$ and is quenched when the electrodes are brought closer. It remains, however, at all times above $1\mu_{\rm B}$ in agreement with earlier theoretical \cite{wang2005,maslyuk_2006} and experimental work \cite{miyajima_2004,miyajima_2007}. Since the benzene rings carry a small negative moment the moment of the molecule as a whole will be reduced below the atomic $3d$ value by about $0.1 - 0.3\mu_{\rm B}$ per benzene ring depending on the electrode separation \cite{wang2005,maslyuk_2006}. The system is expected to have a spin multiplicity of 2 and to be in a $3d^4$ configuration, donating $0.5e^{-}$ to each benzene ring \cite{andrews_1986_2}.  Figures~\ref{dft-results-v}(a) and (d) show the LSDA density of states and the transmission functions for different $d$. One can see the increasing polarization inside the $A_1$ channel, also increasing, but still smaller in the other two channels. This 
is also reflected in the transmission that especially in the vicinity of the Fermi level shows increasing spin polarization, since it stems mostly from the $A_1$ channel. Depending on the bias voltage the current through the VBz$_2$ will be of one spin species or the other exclusively, making this system interesting in the field of spintronics \cite{xiang2006,maslyuk_2006}. At $d=3.8$~\r{A}, for example, the system can be characterized as a half-metal since the density of states is metallic at the Fermi level for the minority electrons and shows a gap for the majority electrons. Our calculations agree qualitatively with the transport calculations by Maslyuk et al. for V$_3$Bz$_4$ clusters \cite{maslyuk_2006}.

\begin{figure}[t]
  \begin{center}
    \includegraphics[width=0.98\linewidth]{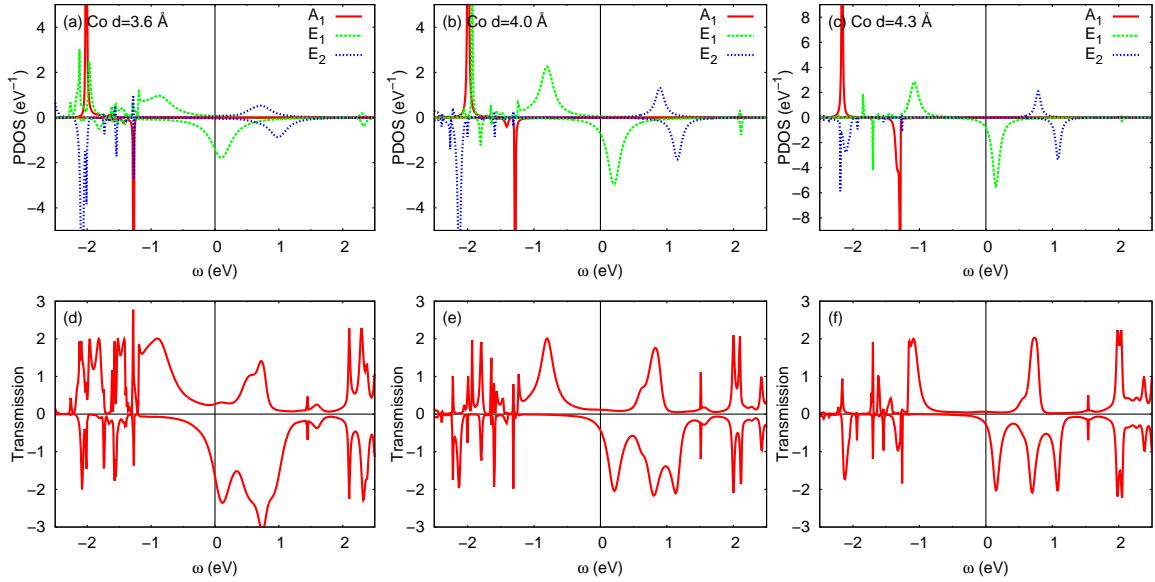}
  \end{center}
  \caption{
    (a) Projected density of states for the $3d$ shell of Co for $d=3.6$~\r{A}, (b) $d=4.0$~\r{A} and (c) $d=4.3$~\r{A}. The corresponsing transmission functions are shown in panels (d), (e) and (f) respectively.
  }
  \label{dft-results-co}
\end{figure}

CoBz$_2$ follows in the same direction with an increasing magnetic moment at increased electrode separation. The states at and in the vicinity of the Fermi level are now dominated by the $E_1$ and $E_2$ channels, the $A_1$ orbital being full. The magnetic moment in LSDA is quenched at small distances, while in B3LYP it constantly remains at about $2\mu_{\rm B}$. Also the total filling increases the more the molecule decouples from the leads. Figure~\ref{dft-results-co} shows the LSDA density of states and the transmission function for different $d$. The $A_1$ channel is also spin polarized and shows large spectral weight at 1 - 2.5eV below the Fermi energy. Calculations for the free molecule in symmetric or asymmetric sandwich geometry show a magnetic moment of about $1\mu_{\rm B}$ for the whole molecule and about $1.4\mu_{\rm B}$ for the $3d$ orbitals only. This is in agreement with earlier calculations by Zhang et al. for the free asymmetric sandwich structure \cite{zhang_2008}. So again the enhanced 
interaction with the benzene rings and the leads considerably changes the magnetic properties of the molecule. The transmission shows some smooth dependence on molecule-lead distance, with increasing sharp molecular resonances at larger $d$. One can also, as for VBz$_2$ identify regions where the density of states and transmission is fully spin polarized, e.g. for $d=4.0$~\r{A} and $d=4.3$~\r{A} at small positive biases, making spintronics applications conceivable. Since the larger clusters of Co$_n$Bz$_m$ form closed rice-ball like clusters \cite{kurikawa1999} growing wires as in the case of V$_n$Bz$_m$ is probably not possible, limiting possible applications in this direction to the single sandwich.

\begin{figure}[t]
  \begin{center}
    \includegraphics[width=0.98\linewidth]{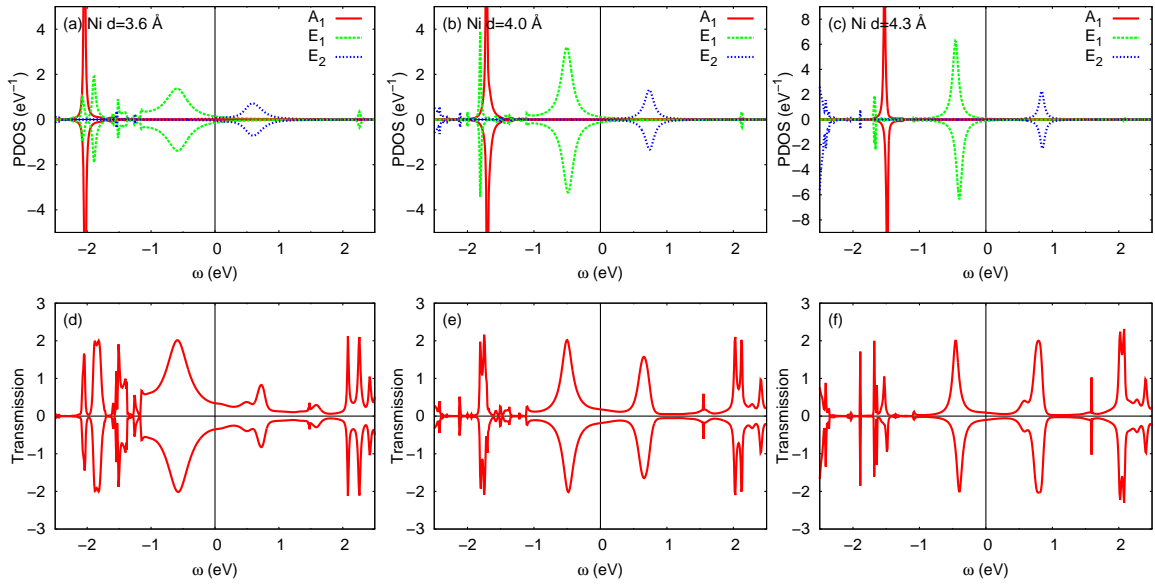}
  \end{center}
  \caption{
    (a) Projected density of states for the $3d$ shell of Ni for $d=3.6$~\r{A}, (b) $d=4.0$~\r{A} and (c) $d=4.3$~\r{A}. The corresponsing transmission functions are shown in panels (d), (e) and (f) respectively.
  }
  \label{dft-results-ni}
\end{figure}

Albeit being very similar in structure to CoBz$_2$ the Ni centered molecule behaves differently. Its projected density of states shows the logical evolution of the foregoing molecule with one more electron in the $3d$ shell. Similarly as in Co the states at the Fermi level are the $E_1$ and $E_2$ states while the $A_1$ shell is almost full showing spectral weight at energies 2eV below the Fermi level, see Fig. \ref{dft-results-ni}(c). The Ni atom is not magnetic, the asymmetry between spin projections being negligible. This is in line with measurements and calculations of NiBz$_2$ \cite{rao_2002} and also typical for Ni adatoms on gold and silver surfaces \cite{beckmann_1996,lazarovits_2002,sandra2012}, see also the next chapter. Accordingly, the NiBz$_2$ sandwich is never magnetic. The transmission is of course also fully symmetric in both spin channels, prohibiting possible spintronics applications in this case.

So we find in general three classes of molecules, depending on the valence of the transition metal center: 1.) Nonmagnetic molecules (ScBz$_2$,NiBz$_2$) 2.) Molecules whose moment is enhanced when compressed by electrodes (TiBz$_2$) and 3.) Molecules whose moment is quenched when compressed by electrodes (VBz$_2$,CoBz$_2$).

\section{Anderson Model Treatment}

\subsection{Discussion of LDA Hybridization Functions}

First, we discuss some general features of the hybridization functions applying to all molecules in the light of an Anderson model treatment. In general the molecules can be compressed considerably by reducing the molecule-lead distance as shown in Fig.~\ref{general1}(b). When the 
leads are sufficiently far away, the molecules relax towards their equilibrium distances which are in agreement with the ones 
computed in Ref. \cite{pandey_2001}. The differences in size for different TM centers can be explained by the 
occupation of bonding or anti-bonding symmetry adapted orbitals \cite{pandey_2001}.

\begin{figure}[t]
  \begin{center}
    \includegraphics[width=0.98\linewidth]{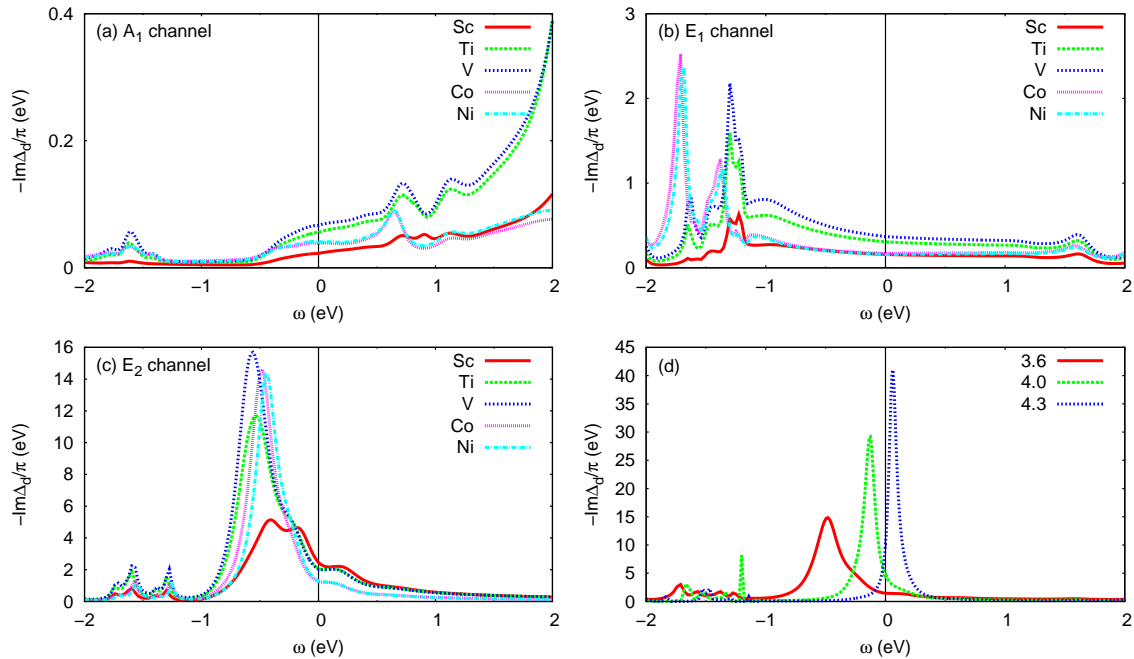}
  \end{center}
  \caption{(a) Imaginary part of the hybridization function for the $A_1$-channel for different TM centers.
    For Ni and Co the distance between lead and TM center is $d=3.6$~\r{A} and for Sc, Ti and V $d=3.2$~\r{A}.
    (b) Same as (a) but for the $E_1$-channel.
    (c) Same as (c) but for the $E_2$-channel.
    (d) Total imaginary part of hybridization function for Co center (summed for all Co $3d$-orbitals) 
    in dependence on the distance between lead and Co center.
  }
  \label{dft-results}
\end{figure}

Figures~\ref{dft-results}(a)-(c) show the imaginary parts of the hybridization functions obtained from the LDA electronic structure 
for different TM centers. Obviously, the general structure of the hybridization function is generic for all molecules:
The imaginary part of the hybridization function exhibits a distinct peak close to the Fermi level 
($E_{\rm F}$) in the $E_2$ channel, whose position, width and height depend significantly on the molecular geometry, 
specifically on the Bz-TM distance as shown in Fig. \ref{dft-results} (d) for CoBz$_2$. Close to $E_{\rm F}$ the $E_1$ channel shows a small hybridization and that of the $A_1$ channel is negligible. 
The hybridization in general, however, increases when the molecule is compressed by the leads. This increase at smaller $h$ could bring 
about a Kondo effect in the $E_1$ or the $E_2$ channels, as we have shown for CoBz$_2$ \cite{sandwich_prl}. 
The dominant feature in the hybridization in the $E_2$ channel stems, similarly as shown for graphene \cite{wehling2010,jacob2010_1} 
from hybridization with the $\pi_z$ molecular orbital state of the benzene rings. The feature does not depend qualitatively on the DFT functional 
used, as we have found the same feature within GGA and also in B3LYP calculations. The $E_1$ orbitals show some interaction with the $\pi$ system of the Bz rings, but relatively far away from the Fermi level. Close to the Fermi level the magnitude of the hybridization is quite low compared to the $E_2$ channel. On the other hand, the $A_1$ orbitals do not hybridize with the Benzene rings for symmetry reasons as we already mentioned above 
(again similar to the case of Co on graphene \cite{wehling2010,jacob2010_1}). Hence these orbitals only couple directly 
to the conduction electrons in the leads explaining their small to negligible hybridizations. The presence of strong molecular resonances 
in the hybridization function makes this case different from the case of 
nanocontacts with magnetic impurities where the hybridization functions are generally 
much smoother (see Ref. \cite{jacob2009} for comparison).

At larger distances between the molecule and the leads, the molecular character of the sandwich becomes more pronounced. 
This is reflected in the hybridization function as shown in Fig.~\ref{dft-results}(d) for the CoBz$_2$ sandwich molecule: 
For larger distances the peak in the hybridization function corresponding to the $E_2$ channel coupling to the $\pi$-molecular
orbital of the benzene rings becomes sharper and shifts to higher energies. This strong dependence of the hybridization in
the $E_2$ channel on the molecular geometry gives us a handle for controlling the electronic structure and 
in particular the Kondo effect by compressing or stretching the molecule with the Cu contacts.

\subsection{Orbital Kondo Effect in CoBz$_{\mathbf{2}}$ and VBz$_{\mathbf{2}}$}

\begin{figure}[t]
  \begin{center}
    \includegraphics[width=0.98\linewidth]{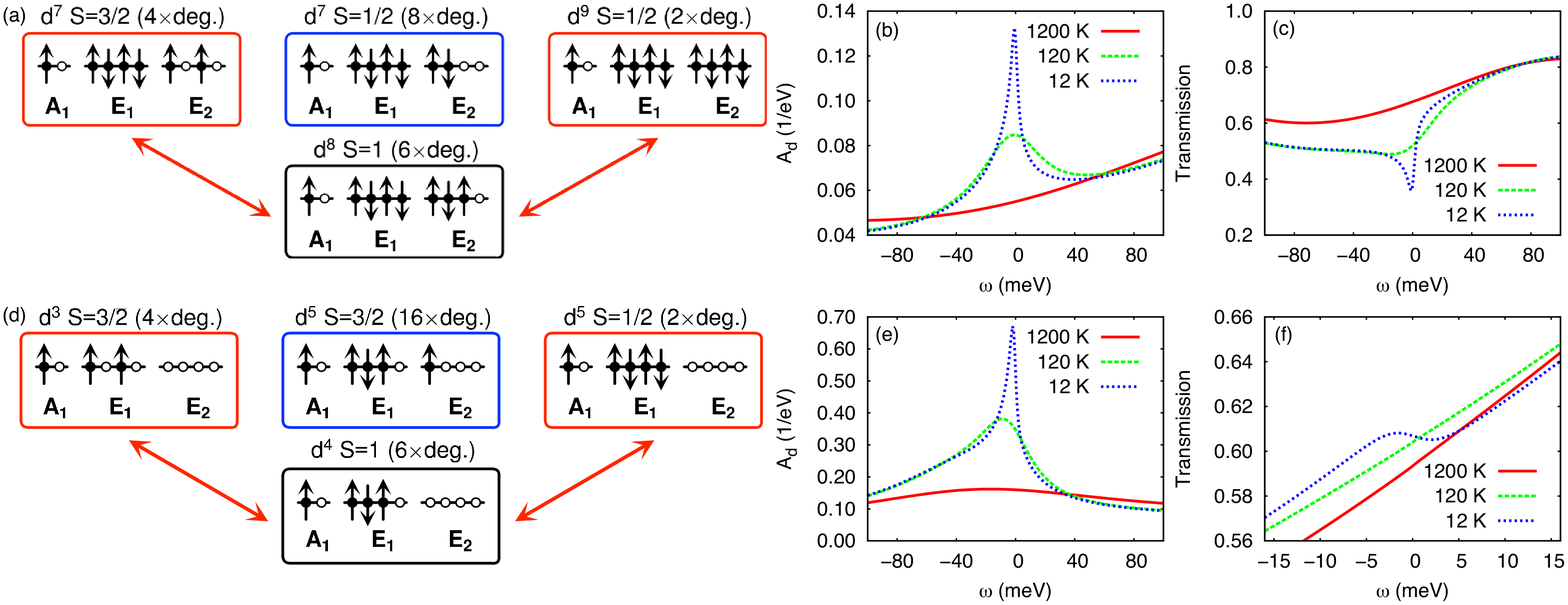}
  \end{center}
  \caption{(a,d)Atomic states governing the low energy physics of the molecules for the smallest distance used, for CoBz$_2$ and VBz$_{2}$, respectively. The dominant contribution is shown lowest with a black frame. The fluctuations bringing about the orbital Kondo effect are indicated by a red frame and arrows along with another important non-Kondo contribution. (b,e) Low energy spectral functions (c,f) low energy transmission functions, for CoBz$_2$ and VBz$_{2}$, respectively.}
  \label{kondo-kondo-kondo}
\end{figure}

We have performed DFT+OCA calculations for the whole series of molecules, the results of this complete study are, however, reported elsewhere \cite{mk_unpub}. Here we instead focus on the low energy spectra and transmission functions of the molecules that we have found to exhibit a Kondo effect: CoBz$_2$ and VBz$_2$. The strong crystal field combined with the Coulomb interaction modify the electronic structure considerably. We begin with CoBz$_2$, which we discussed already in another paper of ours \cite{sandwich_prl} and include it here for completeness. Here we applied the interaction parameters U$=5$eV, J$=1$eV. The only channel showing considerable weight at the Fermi level is the $E_2$ channel, which retains a filling of about 2.7$e^-$ (2.7 electrons), very close to its LDA value, meaning that it holds 3 electrons and a $S=1/2$ predominantly. 
In the other channels the distribution changes: The $A_1$ channel looses one electron as compared to LDA and is now half filled, while the $E_1$ channel gains one electron to be almost full at 3.75$e^-$. This leaves the system in a (predominantly) $S=1$ state, with one spin $1/2$ in the $A_1$ channel and one in the $E_2$. Most importantly, for $d$ around 3.6~\r{A} when the molecule is slightly compressed, 
a sharp temperature-dependent peak appears right at $E_{\rm F}$, as can be seen 
from Fig. \ref{kondo-kondo-kondo}(b). The peak is strongly renormalized (i.e. it only carries 
a small fraction of the spectral weight) due to the strong electron-electron interactions. 

The sharp peak in the spectral function at $E_{\rm F}$ that starts to develop already 
at temperatures of $k_{\rm B}T=0.01$~eV$\approx120$~K stems from the $E_2$ channel which is the channel with the strongest hybridization near $E_{\rm F}$, see Fig.~\ref{dft-results}(c), cf. the discussion of the hybridizations in the previous section.  
Correspondingly, the transmission function (Fig. \ref{kondo-kondo-kondo}(c)) shows a Fano-like 
feature around zero energy. A renormalized, sharp and temperature-dependent resonance in 
the spectral function at $E_{\rm F}$ is commonly associated with the Kondo effect \cite{hewson}. Looking at 
the orbital occupations, we find that the $E_2$ channel that gives rise to the resonance 
for $d=3.6$~\r{A} has an occupation of about $2.8e^-$ while the total occupation of the 
$3d$ shell is $N_{\rm 3d}\sim7.5e^-$ as shown in table \ref{lda-fillings}. The fractional occupation numbers indicate the presence of valence fluctuations where the charges in the individual impurity levels fluctuate in contrast to the \textit{pure}, $s-d$ model-like, Kondo regime, where these fluctuations are frozen \cite{hewson}.
Since the OCA solver allows for the analysis of the contributions of atomic states to the many-body wave function, similarly to the sector analysis on CT-QMC we can use this information to get more information about the possible Kondo state and its origin. Analyzing the atomic states of the Co $3d$ shell contributing to the ground state of the system we find that the principal contribution ($\sim45$\%) is an atomic state with $8$ electrons and a total spin of $S=1$ $(d^8, S=1)$ as shown in Fig.~\ref{kondo-kondo-kondo}(a) (upper panels). The total spin $1$ stems from holes in the $E_2$ and $A_1$ channels. The charge fluctuations in the $E_2$ channel are mainly due to the contribution ($\sim17$\%) of an atomic $(d^7, S=3/2)$ state. There are considerably weaker contributions ($\sim4$\%) from atomic $(d^7, S=1/2)$ and $(d^9, S=1/2)$ states. The individual contributions of the remaining 
atomic states are very small (below $1$\%) but add up to a total contribution of 34\%. 

By exclusion of individual atomic states from the calculation of the spectra we can determine which fluctuations are responsible for the different spectral features. We find that the fluctuations between the $(d^8,S=1)$ and the $(d^7,S=3/2)$ states are primarily responsible for the spectral features close to $E_{\rm F}$ including the sharp Kondo peak right at $E_{\rm F}$.

Note that the fluctuations from the $(d^8,S=1)$ to the $(d^7,S=3/2)$ states that give rise to the Kondo peak at $E_{\rm F}$ actually cannot lead to a spin Kondo effect since the spin $3/2$ of the $d^7$ state is higher than the spin $1$ of the principal $d^8$ state. This means that the spin cannot be flipped via virtual fluctuations involving the $(d^7,S=3/2)$ state.
Instead the fluctuations between the $(d^8,S=1)$ and the $(d^7,S=3/2)$ states which give rise to the Kondo resonance at $E_{\rm F}$ correspond to an {\it orbital Kondo effect} in the doubly-degenerate $E_2$ levels of the Co $3d$ shell. Here the index labeling the two orbitals with $E_2$ symmetry takes the role of a pseudo spin. One can for example assign $\ket{+}$ to one orbital and $\ket{-}$ to the other akin to the physical spin $\frac{1}{2}$. In the principal $d^8$ atomic state the $E_2$ levels are occupied with three electrons and hence have a pseudo spin of $1/2$. By excitation to the $(d^7,S=3/2)$ state the electron with minority real spin and with some pseudo spin state is annihilated. By relaxation to the principal electronic $(d^8,S=1)$ state a minority real spin electron can now be created in one of the two pseudo spin states. Those processes that lead to a flip of the pseudo spin then give rise to the orbital Kondo effect and the formation of the Kondo peak at $E_{\rm F}$.

The absence of a spin Kondo effect where the total spin $1$ of the principal $d^8$ atomic state is screened, is understood on the following grounds: First, in general the Kondo scale decreases exponentially with increasing spin of the magnetic impurity \cite{nevidomskyy_2009}. In addition, here the $A_1$ level does not couple at all to the conduction electrons around $E_{\rm F}$ (no hybridization). Thus the spin $1/2$ associated
with it cannot be flipped directly through hopping processes with the conduction electron bath.  

On the other hand, an {\it underscreened} Kondo effect as reported in Ref. \cite{parks2010} where only the spin $1/2$ within the $E_2$ shell is screened is also suppressed compared to the orbital Kondo effect due to Hund's rule coupling: Screening of the spin $1/2$ in the $E_2$ shell
can take place by fluctuations to the ($d^7,S=1/2$) state. However, the Hund's rule coupling $J$ favors the high spin ($d^7,S=3/2$) state over the low spin ($d^7,S=1/2$) state as can also be seen from the smaller weight of the latter compared to the former.
Hence the Kondo scale is considerably lower for the underscreened Kondo effect than for the orbital Kondo effect found here. At even lower temperatures (not accessible within the OCA) the two Kondo effects may in fact coexist. Hence in principle the setup holds the possibility of an SU(4) Kondo effect \cite{borda_2003}, as also observed in other molecular systems \cite{novel_kondo}.

We have checked the dependence of the LDA+OCA spectra on the double counting correction (DCC) as well as on the interaction parameters $U$ and $J$. We find that the Kondo peak is qualitatively stable for $U$ ranging from 3 to 7~eV, 
and for $J$ ranging from 0.5 to 1~eV. The physics of the system is also qualitatively 
stable against variations of the DCC of a few eV around the fully localized limit (FLL) correction.
The contributions of atomic states to the ground state show a continuous trend while the parameters are varied. The largest contribution to the ground states still stems from the ($d^8,S=1$) state. Depending on the choice of parameters the size of the $d^7$ and $d^9$ contributions is varying. In particular, changes in the DCC (and consequently the filling of the Co $3d$ shell) 
shift the balance between $d^7$ and $d^9$ admixtures to the ground state.  
So in general we find the spectra and also the Kondo peak to be 
qualitatively robust against shifts of the impurity levels in energy
over a range of several electron volts, and changes of $U$ between 3 
and $7$~eV, and of $J$ between $0.5$ to $1$~eV. As expected for the Kondo effect
the sharp resonance stays pinned to the Fermi level when shifting the 
impurity levels in energy, and only height and width somewhat change. The stability of the Kondo effect against variations of $J$ is another hint towards our conjecture that we have found an effect related to orbital degrees of freedom and $\textit{not}$ to spin. For the spin Kondo effect the dependence of the Kondo temperature on $J$ is exponential \cite{hewson}, and thus we would expect a strong influence of a variation in $J$, which is not the case here.

Stretching the molecule by displacing the tips of the Cu nanowires the Kondo resonance and the concomitant Fano line shape in the 
transmission disappear for distances $d\ge4$~\r{A}. This is accompanied 
by an increase of the occupation of the Co $3d$ shell, see table \ref{lda-fillings}. The new regime is characterized by a strong valence mixing between the $d^8$ and $d^9$ atomic state of roughly equal contribution indicating that the system is now in the so-called mixed valence regime (see e.g. Ref.~\cite{hewson}, Chap. 5). Hence the orbital Kondo effect and the associated spectral features can be controlled by stretching or compressing the molecule via the tip atoms of the Cu nanocontact. This strong sensitivity on the molecular conformation stems from the sharp features in the hybridization function which change considerably when the molecule is stretched or compressed as can be seen from 
Fig.~\ref{dft-results}(d). This peculiar behavior is qualitatively different from the case of the nanocontacts containing magnetic 
impurities where the hybridization functions are much smoother \cite{jacob2009}. 

Let us now turn to VBz$_2$. Here we employed the parameters U$=3.5$eV, J$=0.75$eV. Figure~\ref{dft-results-v}(a) and (d) show the LSDA density of states at $d=3.2$~\r{A} and the transmission function for different $d$. In contrast to the case of Co the density of states at the Fermi level, as well as the spin, is dominated by the $A_1$ channel. The spin polarization is still small, the $E_1$ and $E_2$ channels showing no spin polarization at all. The $d$ shell is in total filled with about 4$e^-$ in LSDA as well as LDA, see Tab.~\ref{lda-fillings}. 

Since the total charge was fixed close to its LDA value of 4$e^-$, for reasons discussed above in the methodology section, only a redistribution of charge was possible within the $3d$ shell. This redistribution is, however, considerable and from table \ref{lda-fillings} we can identify two regimes. In the first regime at $d=3.2$~\r{A} the $E_1$ channel, formerly occupied by one electron increases its filling to about $2.6e^-$ at the expense of the other channels, that are now occupied by less than one electron each. This means, similarly to the Co $E_2$ shell discussed above, the V $E_1$ shell now predominantly holds 3 electrons and a $S=1/2$, while the remaining channels hold about one electron altogether. In the second regime at $d > 3.2$~\r{A} the charge redistributes such, that the $A_1$ and $E_2$ channels hold one electron each, while the $E_1$ channel is occupied by two electrons.

However, let us for the moment remain at $d=3.2$~\r{A} since in this case a sharp temperature-dependent peak appears at $E_{\rm F}$, as shown 
in Fig. \ref{kondo-kondo-kondo}(b). Its formation can be observed at temperatures below $k_{\rm B}T=0.005\mathrm{eV}\approx120\mathrm{K}$. The resonance develops only in the $E_1$ channel and only at $d=3.2$~\r{A} in the charge regime discussed above. An analysis of the hybridization function shown in Fig.~\ref{dft-results} indicates that the molecule has been compressed to such extent, that the hybridization in the $E_1$ channel has risen considerably; higher than for any other molecule we considered. 

In this case the $3d$ shell is filled with $\sim 4e^-$, distributed to the different channels as follows: $A_1: 0.9e^-$, $E_1: 2.6e^-$ and $E_2: 0.6e^-$.
By exclusion of individual atomic states from the calculation of the spectral functions we can determine which fluctuations are responsible for the different spectral features.  In the case of V there is one atomic $(d^4, S=1)$ state that contributes with $\sim30$\% to the ground state, see Fig. \ref{kondo-kondo-kondo}(a) (lower panel). Other notable contributions stem from a $(d^3, S=3/2)$, a $(d^5, S=3/2)$ and a $(d^5, S=1/2)$ state. All these four atomic states have one electron in the $A_1$ channel and an empty $E_2$ channel.

Our analysis indicates that an orbital Kondo effect now occurs in the $E_1$ channel. In this case the orbital degree of freedom in the doubly degenerate $E_1$ channel takes the role of a pseudo-spin, which is then screened by the conduction electrons of the leads. Here, we can again exclude that the spin degree of freedom is the most important one. The reasoning here is completely analogous to the Kondo effect in the $E_2$ channel of CoBz$_2$. One possibility for a spin Kondo effect would be a spin $1$ Kondo effect screening the total spin in the $3d$ shell. This is possible, but very improbable since the $A_1$ channel couples to the bath very weakly, thus a flipping of the spin via exchange of particles with the bath is very unlikely. A second possibility would be an underscreened spin $S=1/2$ Kondo effect occurring only in the $E_1$ shell. This is again possible, but is strongly suppressed due to the Hund's coupling to the $A_1$ electron. Additionally, the states bringing about the resonance, shown in Fig. 
\ref{kondo-kondo-kondo}(a) (lower panel), are not the ones required for a spin Kondo effect. Thus, the only effect that is consistent with our observations is the orbital Kondo effect in the $E_1$ channel. 
 
Correspondingly, the transmission function shows a small feature at the Fermi level. The size of this feature is much smaller than seen for CoBz$_2$, since the $E_1$ channel has much smaller hybridization with the rest of the system due to symmetry as discussed above. Therefore the indirect effects of the local correlations on the surroundings like the feature in the transmission are much smaller also \cite{frank_2015}.

\section{Conclusions}

We have investigated the TMBz$_2$ molecules with Sc,Ti,V,Co and Ni centers coupled to Cu 
nanowires. We have computed how the structural parameters as well as the electronic and magnetic properties change when the molecules are coupled to Cu electrodes in a nanocontact geometry. We identify potential candidates for half-metallic behaviour and spintronics applications.
Our study shows, that the parameters of a generalized Anderson model can be tuned using a real material. In the case of TMBz$_2$ molecules in a nanocontact the hybridization strenght ($\Delta$) can be controlled by the proximity of the electrodes to the molecule, while the local Coulomb interaction ($U,J$) as well as the filling of the $3d$ shell ($n$) are controlled by species of the central atom. 
In this way a direct connection between a simple theoretical model and a realistic experimental setup can be made. 
In addition we have identified two molecules, CoBz$_2$ and VBz$_2$, that in our model exhibit the unusual orbital Kondo effect. The Kondo effect occurs in the doubly degenerate $E_1$ or $E_2$ shells in vanadium and cobalt respectively and follows the same mechanism in both cases. Further studies might be interesting for larger instances of the TM$_n$Bz$_m$ class of systems. Nanowires made from the sandwich molecules can become important in spintronics applications. It would thus be worthwhile to investigate how the predictions concerning the properties of such wires, obtained mostly from DFT and other single particle methods, hold up when electronic correlations are included.

\ack
Support from the DFG via FOR1162 and ETSF are acknowledged. 
MK gratefully acknowledges the hospitality of the Max-Planck-Institute 
in Halle (Saale).

\section*{References}

\bibliography{thebib}

\end{document}